\documentclass[Physsubmission, Phys]{SciPost}

\hypersetup{
    colorlinks,
    linkcolor={red!50!black},
    citecolor={blue!50!black},
    urlcolor={blue!80!black}
}

\usepackage[bitstream-charter]{mathdesign}
\DeclareSymbolFont{usualmathcal}{OMS}{cmsy}{m}{n}
\DeclareSymbolFontAlphabet{\mathcal}{usualmathcal}

\begin{document}

\begin{center}{\Large \textbf{
Nuclear matter effects on jet production at electron-ion colliders\\
}}\end{center}

\begin{center}
Hai Tao Li\textsuperscript{1,2},
Ze Long Liu\textsuperscript{3} and
Ivan Vitev\textsuperscript{3$\star$}
\end{center}

\begin{center}
{\bf 1} HEP Division, Argonne National Laboratory, Argonne, Illinois 60439, USA
\\
{\bf 2} Dept. of Physics and Astronomy, Northwestern University, Evanston, Illinois 60208, USA
\\
{\bf 3} Theoretical Division, Los Alamos National Laboratory, Los Alamos, NM, 87545, USA 
\\
* ivitev@lanl.gov 
\end{center}

\begin{center}
\today
\end{center}


\definecolor{palegray}{gray}{0.95}
\begin{center}
\colorbox{palegray}{
  \begin{tabular}{rr}
  \begin{minipage}{0.1\textwidth}
    \includegraphics[width=22mm]{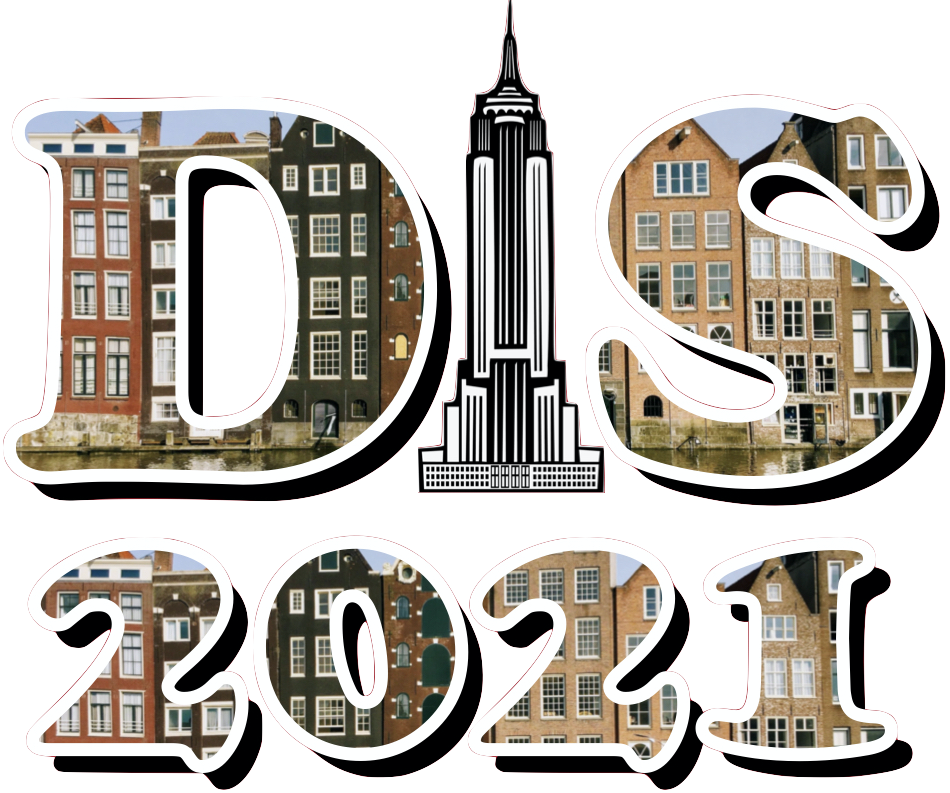}
  \end{minipage}
  &
  \begin{minipage}{0.75\textwidth}
    \begin{center}
    {\it Proceedings for the XXVIII International Workshop\\ on Deep-Inelastic Scattering and
Related Subjects,}\\
    {\it Stony Brook University, New York, USA, 12-16 April 2021} \\
    \doi{10.21468/SciPostPhysProc.?}\\
    \end{center}
  \end{minipage}
\end{tabular}
}
\end{center}

\section*{Abstract}
{\bf
In these proceedings we report recent progress on understanding hadron and jet production in electron-nucleus collisions at the future Electron-Ion Collider \cite{Li:2020zbk,Li:2020rqj}. 
These processes will play an essential role in the exploration of the partonic structure of nuclei and the study  of parton shower  evolution in strongly-interacting matter. We employ the framework of soft-collinear effective theory, generalized to include in-medium interactions, to present the first theoretical results for inclusive hadron and jet cross sections, as well as the jet charge modification in  deep inelastic scattering  on nuclei. We further demonstrate how to separate initial-state and final-state effects. 
 }


\section{Introduction}
\label{sec:intro}

The future high-luminosity and high-energy Electron-Ion collider (EIC) will help answer questions central to nuclear science. These include the 3D structure of the nucleon, the origin of mass, the  properties of the high density gluon saturation regime, and the physics of hadronization in strongly-interacting matter~\cite{AbdulKhalek:2021gbh}. Among these topics the physics of light and heavy  hadron and jet production in 
 deep inelastic scattering  (DIS) on nuclei is, perhaps,  the least developed. In these proceedings we report  the  first calculations of inclusive hadron and  jet production and the jet charge in electron-nucleus collisions at the EIC and investigate the impact of initial-state and final-state cold nuclear matter effects. To develop  the electron-nucleus  (e+A)   jet physics program at the EIC,  we find it useful to place emphasis on  observables that have been the most illuminating in the case of heavy-ion collisions.  At the same time, the kinematics  in DIS is very different relative when compared to  heavy-ion collisions (A+A).  In the latter case the hard parton and the associated shower propagate in a quark-gluon plasma that is,  to first approximation, static in the frame where the jet momentum is transverse to the collision axis.   On the other hand, at the EIC even at forward  rapidity in the proton / nucleus going direction the energy of the jet is dominated by the longitudinal momentum component in the rest frame of the nucleus.  It is important to  establish a theoretical approach for calculating jet and hadron production and to be able to predict the magnitude of nuclear modification relative to the electron-proton (e+p) case.  Furthermore, to address the physics of  parton and particle transport in matter and the physics of gluon saturation, it is essential to be able to separate initial-state and final-state effects.  We have developed strategies to accomplish this task and discuss them here. Quenching of jet cross sections in cold nuclear matter is complemented by a modification of the jet substructure.  
We illustrate this using the jet charge observable. Finally, inclusive jet measurements at the EIC will go in parallel with studies of heavy flavor-tagged jets. The theory developments that we report here can be generalized to charm quark jets and bottom quark jets to shed light on heavy quark mass effects on parton shower formation.

\section{Theoretical formalism}
\label{sec:formalism}

In QCD factorization, the inclusive hadron and jet cross sections can be expressed as follows \cite{Li:2020zbk,Li:2020rqj}, 
 \begin{equation}
E_{h/J} \frac{d^{3} \sigma^{l N \rightarrow h/J X}}{d^{3} P_{j/J}} =\frac{1}{S} \sum_{i, f} \int_{0}^{1} \frac{d x}{x} \int_{0}^{1} \frac{d z}{z^{2}} f_{i / N}(x, \mu) 
  \hat{\sigma}^{i \rightarrow f}(s, t, u, \mu) O_{f }(z, \mu) \;, 
\label{fact}
\end{equation}
where $ f_{i / N}$ is the PDF of parton $i$ in nucleon $N$.  $\hat{\sigma}^{i\to f}$ is the partonic cross section with initial-state parton $i$  and final-state parton $f$, which we take  up to NLO and account for the resolved photon contribution.  For hadron production 
 $O_{f }(z, \mu)  \equiv    D_{h/f }(z, \mu)$ is the fragmentation function,  and for jet production   $O_{f }(z, \mu)  \equiv   J_{f }(z,  p_TR, \mu)$   is   the  semi-inclusive jet functions (SiJFs) initialed by parton $f$.
When the jet radius $R$ is small,  potentially large logarithms of the type $\ln R$ are resummed via time-like DGLAP evolution.

 Jet substructure on the other hand is sensitive to  the radiation pattern inside a given jet and is governed by a smaller intrinsic scale.  
One such observable is the average jet charge defined as the transverse momentum $p_T^{i}$  weighted sum of the charges $Q_i $ of the jet constituents,
\begin{equation} \label{eq:charge}
    Q_{\kappa, {\rm jet}}  = \frac{1}{\left(p_T^{\rm jet}\right)^\kappa } \sum_{\rm i\in jet} Q_i \left(p_T^{i} \right)^{\kappa } \; ,  \quad \kappa > 0 \; ,
\end{equation}
where $\kappa$ is a free parameter that must be positive for infrared safety. The jet charge is strongly correlated with the electric charge of the  parent parton and can 
be used to separate  quark jets from anti-quark jets and to determine their flavor origin.

In reactions with nuclei the factorization formula Eq.~(\ref{fact}) receives in-medium corrections~\cite{Li:2020zbk,Li:2020rqj}. If the leading-twist distributions of partons in nuclei are modified relative to the nucleon, they can be accounted for
through nuclear PDFs. Our main focus,  however,  is on the effect of final-state interactions.   The full  fragmentation function evolution in the presence of nuclear matter is  given by
\begin{equation} \label{eq:fullevol}
    \frac{d}{d \ln \mu^{2}} {D}^{h/i}\left(x, \mu\right)= 
    \sum_{j} \int_{x}^{1} \frac{d z}{z}  {D}^{h/j}\left(\frac{x}{z}, \mu\right)  
   \left( P_{j i}\left(z, \alpha_{s}\left(\mu\right)\right) +  P_{j i}^{\rm med}\left(z, \mu\right)  \right)   \, ,  
\end{equation}
where in  Eq.~(\ref{eq:fullevol})  $ P_{j i}^{\rm med}$ are the medium corrections to the splitting functions. Their derivation with emphasis on  DIS can be found in~\cite{Sievert:2018imd} and they lead to broader and softer
parton showers than the vacuum ones~\footnote{Higher order corrections do not qualitatively change this picture~\cite{Fickinger:2013xwa}.}.
The SiJFs, on the other hand, receive a medium induced contribution  at NLO that can be written as 
\begin{equation}
J_{f} (z,p_TR, \mu ) = J_{f}^{\rm vac} (z,p_TR, \mu )+  J_{f}^{\rm med}(z,p_TR, \mu )  \, , 
\end{equation}
where the vacuum contributions are calculated at the LL accuracy, while only the fixed-order medium corrections are included consistently.   It is important to note that $J_{f}^{\rm med}(z, p_TR, \mu )$ can be expressed in terms of the in-medium splitting functions~\cite{Li:2020rqj}. The factorization formula for the  jet charge observable Eq.~(\ref{eq:charge})  also receives corrections from final-state interactions. Thus,
we expect to see differences between the substructure of jets in e+p and e+A reactions, although smaller than  in the case of heavy ion reactions.

\section{Selected results}
\label{sec:results}
\begin{figure*}[t!]
	\centering
	\includegraphics[width=0.46\textwidth]{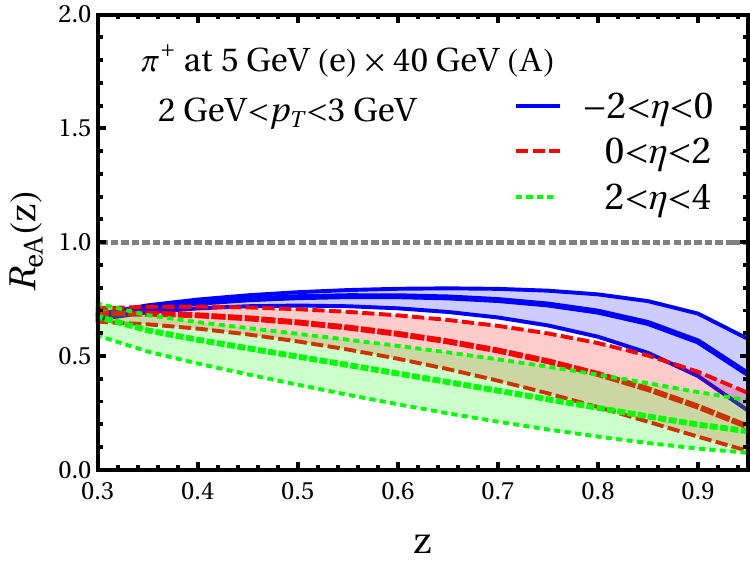}\,\,\,  	\includegraphics[width=0.46\textwidth]{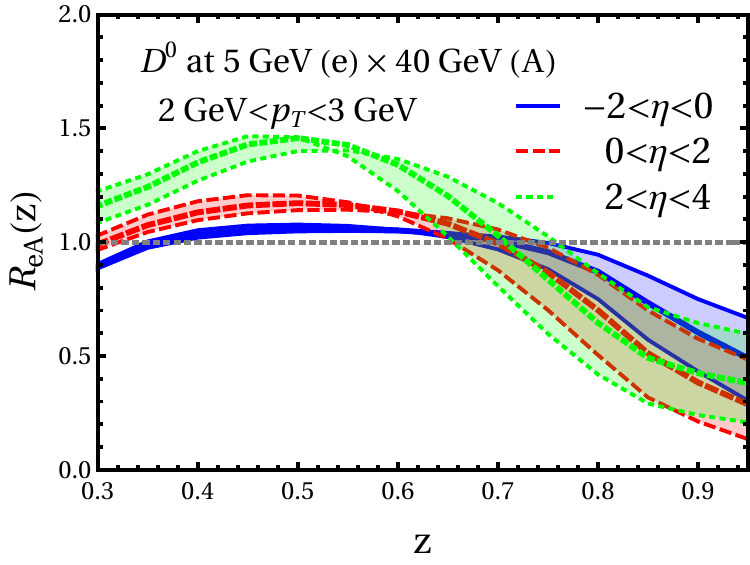}
	\caption{ Left: in-medium corrections for $\pi^+$ production as a function of $z$ at the EIC  in three rapidity regions. 
		Blue bands (solid lines), red bands (dashed lines), and  green bands (dotted lines) correspond to  -2$<\eta<$0, 0$<\eta<$2 and 2$<\eta<$4, respectively.
		Right: the same in-medium corrections, but  for $D^0$  at the EIC in the same rapidity regions. }
	\label{fig:zdisEIC_pi}
\end{figure*}

 To investigate the nuclear medium effects, we study the ratio of the cross sections in electron-gold (e+Au) collision to the one in e+p collision.  When looking at hadron production, we use the cross section of inclusive jet production for normalization that minimizes  the effect of nuclear PDFs,
\begin{equation}\label{eq:defRAatEIC}
R_{eA}^{h}(p_T,\eta,z)={\frac{N^{h}(p_T,\eta,z)}{N^{\rm inc}(p_T,\eta)}\Big|_{\rm e+ Au}} \Bigg/ {\frac{N^{h}(p_T,\eta,z)}{N^{\rm inc}(p_T,\eta)}\Big|_{\rm e+p}} \, .
\end{equation}
The transport properties of cold nuclear matter can be constrained to a certain extent from HERMES data~\cite{Airapetian:2007vu}. Light hadron production in DIS on nuclei at HERMES, however, has not been able to differentiate between competing 
models of  parton energy loss and hadronization in cold nuclear matter, see Ref.~\cite{Accardi:2009qv} for review.  This is one area where the EIC with its electron-nucleus  program and  center-of-mass energies  sufficient to copiously produce D-mesons and B-mesons can  make a major contribution.  We show our results for $R_{eA}^{h}(p_T,\eta,z)$ in 	Fig.~\ref{fig:zdisEIC_pi} for 5 GeV(e) $\times$ 40 GeV(A) collisions where
nuclear effects are most significant. Suppression for pions in three rapidity intervals is shown in the left panel and is largest in the nucleus-going direction, however it is qualitatively similar to the one observed in 
HERMES data.   In contrast to light flavor, the modification  of open heavy flavor in DIS reactions with nuclei  is much more closely related to the  details of hadronization.  The $R_{eA}(z)$ in the right panel of Fig.~\ref{fig:zdisEIC_pi}  is qualitatively consistent with the  effective modification of fragmentation functions  in that there is a significant suppression for large values of $z$, but it quickly evolves to      
 enhancement for $z<0.65$ for $D$-mesons. This is an unambiguous signal of final-state partonic interactions.

\begin{figure}[!t]
    \centering
    \includegraphics[width=0.48\textwidth]{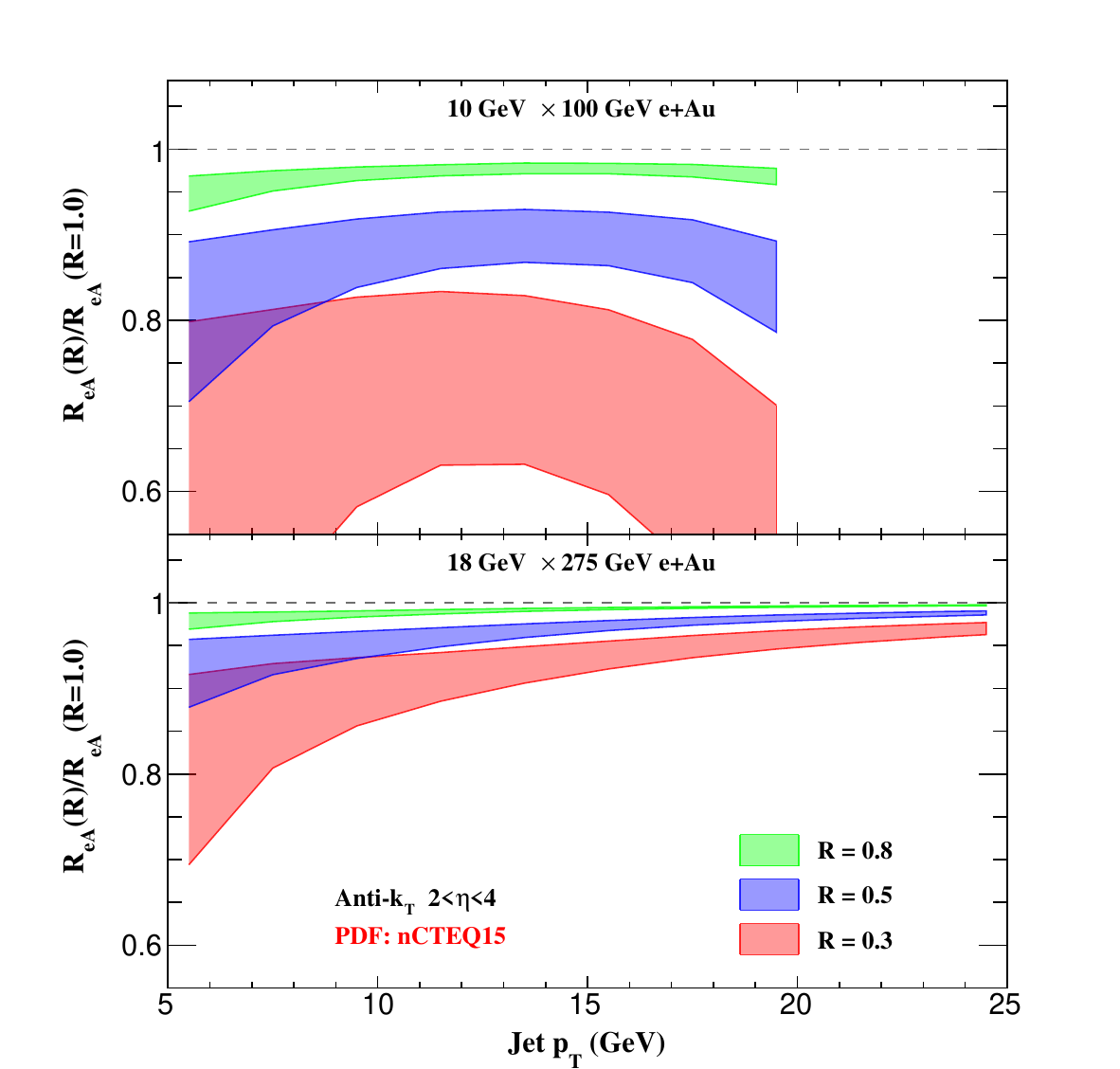}  \  \  \     \includegraphics[width=0.48\textwidth]{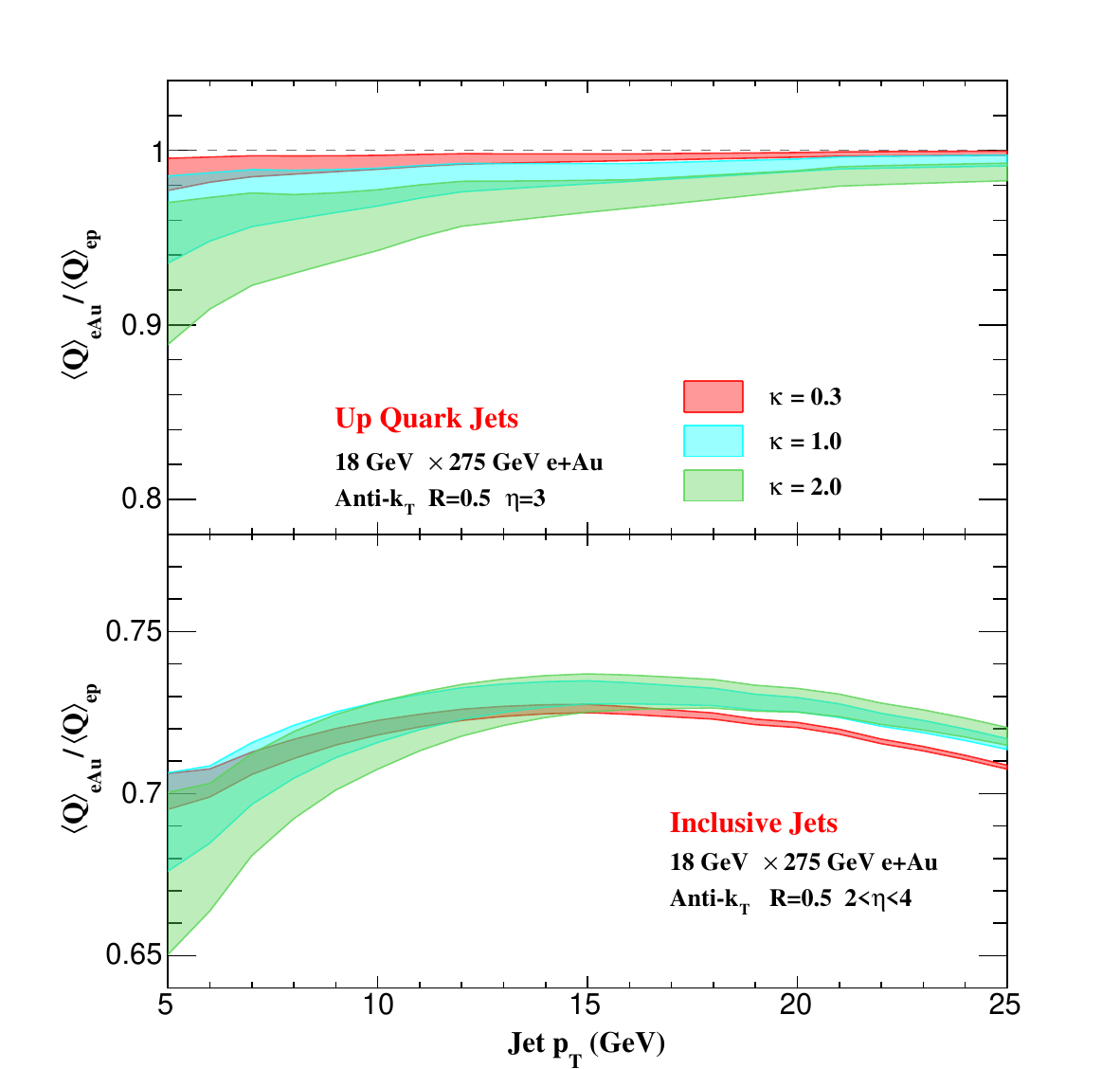} 
    \vspace{-0.cm}
    \caption{Left: ratio of jet  cross section modifications for different  radii $R_{\rm eA}(R)/R_{\rm eA}(R=1.0)$ in 10 $\times$ 100 GeV (upper) and  18 $\times$ 275 GeV (lower) e+Au collisions, where the smaller jet radius is R=0.3, 0.5, and 0.8,  and the jet rapidity  interval is $2<\eta<4$.   Right:  modifications of the jet charge in e+Au collisions. The upper panel is the modification for up-quark jet with $\eta=3$ 
    in the lab frame. The lower panel is the results for inclusive jet with $2<\eta<4$ in 18 $\times$ 275 GeV e+Au collisions.}
    \label{fig:ReAto1}
\end{figure}

For reconstructed jets,  nuclear effects in e+A reacions can be studied through the ratio 
\begin{equation}
    R_{\rm eA}(R) = \frac{1}{A}   {\int_{\eta1}^{\eta2} d\eta \;  d\sigma/d\eta dp_T\bigskip
    |_{e+A}} \,  \Bigg/ \,  {   \int_{\eta1}^{\eta2}  d\eta \;   d\sigma/d\eta dp_T\big|_{e+p}}\,.
\end{equation}
We found that the modification of jets of moderate radius $R=$0.5 is significant,  but mixes initial-state and final-state effects. To isolate  jet quenching effect in matter   we propose to measure the ratio of the suppression  with different jet radii,  $R_{\rm eA}(R)/R_{\rm eA}(R=1)$. Results are shown in the left panel of Fig.~\ref{fig:ReAto1} and the red, blue, and green bands denote ratios with $R=0.3\,,0.5\,,0.8$, respectively. 
The suppression is stronger for smaller center-of-mass energies  and  for 10 GeV(e) $\times$ 100 GeV(A) collisions it is very significant -  close to a factor of two and similar to what has been observed in  heavy-ion collisions. The separation in the magnitude of the effect as a function of $R$ is also very clear, the reason being that medium-induced parton showers are broader than the ones in the vacuum.
In the right panel of  Fig.~\ref{fig:ReAto1}  we present the ratio of the jet charge in e+A relative to the one in e+p collisions. 
The red, blue and green bands correspond to the jet charge parameter $\kappa=0.3\,,1.0\,,2.0$, see Eq.~(\ref{eq:charge}).   If  jets are separated based on their flavor,  for example up quark jets, 
  $\langle Q_{ \kappa,q}^{\rm eA} \rangle/\langle Q_{\kappa,q}^{\rm ep} \rangle$  is a direct measure of medium-induces scaling violations in QCD - of order 10\%.  On the other hand,  the modification of the average charge for inclusive jets behaves differently because there is a  cancellation between contributions from jets initiated by   up quarks and down quarks.  In nuclei the partonic composition is different because of the neutron contribution.  The modification is about 30\% and measurement of the charge for inclusive jets can help constrain isospin effects and  the up/down quark PDFs in the nucleus.

\section{Conclusion}
\label{sec:conclusions}

To summarize, hadron and jet physics in e+p and e+A collisions is an important part of the EIC program. It can provide constraints on nuclear PDFs, the physics of hadronization, and the transport of energy and matter in the nuclear environment. We presented calculations of light hadron and jet cross sections, and also jet substructure in e+p and e+A with transport parameters constrained by HERMES experimental measurements.  Our results show that to access the physics of hadronization and particle propagation in matter, lower CM energies, forward rapidity, and high luminosity will be very beneficial. Heavy flavor is the key to differentiating between models of hadron suppression in large nuclei. To differentially study parton shower formation in matter, we presented results for reconstructed jets.   Using measurements with different jet radii, we  developed strategies to separate initial-state from final-state effects. We further showed, on the example of the  jet charge, that jet substructure is also modified in e+A versus e+p reactions and can be used to study medium-induced scaling violations in QCD.  Upcoming results on charm jets and  bottom jets at the EIC~\cite{HFjet} are also promising. They point to the opportunity for heavy flavor tomography of cold nuclear matter and to novel ways to study the mass hierarchy of jet quenching effects on jet substructures.  It will also be interesting to study sub-eikonal effects on jet observables  related to inhomogeneities in the medium density  and fluctuations in the strength of the underlying gluon fields~\cite{Sadofyev:2021ohn}.

\section*{Acknowledgements}

\paragraph{Funding information}
H.L.,  Z.L., and  I.V. are supported by the  LDRD program at LANL. I.V. is partly supported by the U.S. Department of Energy under Contract No. 89233218CNA000001. 
H.L. is partly supported  by   the U.S. Department of Energy under Contract  No. DE-AC02-06CH11357 and the National Science Foundation under Grant No. NSF-1740142.




\nolinenumbers

\end{document}